\documentclass[prf,aps]{revtex4}
\usepackage{graphicx}
\usepackage{natbib}
\usepackage{color}
\usepackage{amsmath}
\usepackage{enumerate}
\usepackage{hyperref}
\usepackage[toc,page]{appendix}
\def\strutdepth{\dp\strutbox}
\def\nw#1{\strut\vadjust{\kern-\strutdepth\vtop to0pt{\vss\hbox to\hsize
{\hskip\hsize\hskip5pt$\leftarrow$\hss\strut}}}{\em #1}}

\usepackage[makeroom]{cancel}

\begin{document}

\preprint{APS/123-QED}
% **

%\title{Massive reduction in buckling threshold of flexible sheets due to hydrodynamic interactions.}
\title{Hydrodynamic interactions change the buckling threshold of parallel flexible sheets in shear flow}
\author{Hugo Perrin, Heng Li, and Lorenzo Botto}

\affiliation{Process \& Energy Department, Faculty of Mechanical, Maritime and Materials Engineering, Delft Univ. of Technology, Delft, The Netherlands}

\date{\today}

\begin{abstract}

Buckling induced by viscous flow changes the shape of sheet-like nanomaterial particles suspended in liquids. This instability at the particle scale affects collective behavior of suspension flows and has many technological and biological implications. Here, we investigated the effect of viscous hydrodynamic interactions on the morphology of flexible sheets. By analyzing a model experiment using thin sheets suspended in a shear cell, we found that a pair of sheets can bend for a shear rate ten times lower than the buckling threshold defined for a single sheet. This effect is caused by a lateral hydrodynamic force that arises from the disturbance flow field induced by the neighboring sheet. The lateral hydrodynamic force removes the buckling instability but massively enhances the bending deformation. For small separations between sheets, lubrication forces prevail and prevent deformation. Those two opposing effects result in a non-monotonic relation between distances and shear rate for bending. Our study suggests that the morphology of sheet-like particles in suspensions is not purely a material property, but also depends on particle concentration and microstructure.
\end{abstract}

\maketitle
\section{Introduction}
Soft biological or synthetic objects, such as cells, lipid bilayers, macromolecules, and nanoparticles, can deform when suspended in sufficiently strong shear or extensional flows \cite{PhysRevLett123048002,PhysRevLett98188302,D0SM02184A,Marlow2002ve}. Predicting flow-induced morphological changes is crucial in many fields, ranging from biophysics, where swimming of micro-organisms relies on fluid-structure interactions \cite{ANFBacteria}, to soft matter physics, where the rheological response of a particulate suspension is affected by the instantaneous particle shape \cite{oldroy53}. Model studies in canonical flows have provided profound physical insights of general applicability. For example, the theoretical prediction of the coil-stretch transition of polymers in simple shear flow \cite{pgdg74,PhysRevLett108038103} was instrumental in the development of rheological models for dilute polymer solutions \cite{RheoLarson}.

The recent need to develop liquid-based methods to process two-dimensional (2D) nanomaterials \cite{Nicolosi1226419,C3MH00144J,Paton2014ty} has triggered new interest on the effect of flow on the morphology of sheet-like materials \cite{D0SM01630F,D0SM02184A,Nicolosi1226419,C3MH00144J,D0SM02081H,JCP14Green,YuPRF22,Giulapof}. Two-dimensional materials have low bending modulii and therefore can undergo transient or permanent buckling in flow \cite{D0SM02184A}. Recent numerical studies \cite{D0SM02184A,D1SM01510A} demonstrate that purely mechanical models based on the competition between hydrodynamic compressive force and elastic-bending forces can capture the change of morphology of isolated graphene sheet and 2D polymers suspended in a simple shear flow. This agreement demonstrates that the morphology of a single sheet is determined by a buckling instability whose threshold depends, for a given fluid shear rate and  viscosity, only on the bending modulus and length of the sheet. However, the extension of this result to suspensions of many particles is an open question. Because of their relatively large contact area, sheet-like particles are prone to stacking at small inter-particle separations \cite{WANG2020230,C3MH00144J}. Hydrodynamic interactions between nearly parallel sheets are thus expected to alter the buckling dynamics predicted for single sheets.

%%%%%
In this study we investigate parallel pairs of flexible sheets in a shear flow as a function of their separation distance, and study how the buckling instability threshold depends on hydrodynamic interactions. By performing model experiments, and interpreting the results with the help of boundary-integral simulations and theoretical modeling, we demonstrate that hydrodynamic interactions can trigger bending far below the buckling threshold of a single sheet. Hydrodynamic interactions cannot therefore be considered second-order effects when predicting the morphology of flexible sheets in flow. More specifically, our simulations and theoretical modeling show that the dipolar disturbance flow field induced by each sheet gives rise to a lateral hydrodynamic force. This lateral force modifies the mechanical response of the sheet pair to the compressive axial hydrodynamic force experienced when the pair is oriented in the compressional quadrant of the shear flow. On the other hand, for small separations, the lubrication forces overcome this dipolar contribution and prevent bending. These two competing effects result in a non-monotonic relation between interparticle distance and critical shear rate for buckling. More generally, our results suggest that the deformation of close sheets in a suspension may not only depend on the mechanical and geometric properties of each sheet but also strongly on the pair-particle separation and thus on the concentration.
%%%%%
\section{ Experiments}
\begin{figure}[h]
\begin{center}
\includegraphics{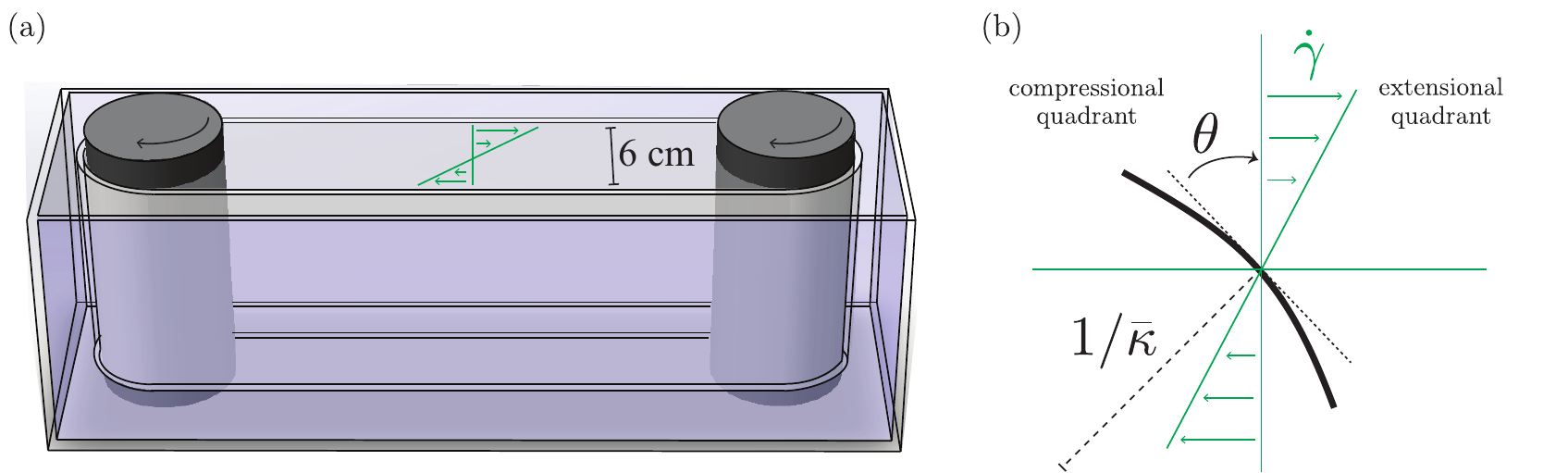}
\caption{(a) Schematic of the shear cell. (b) Schematic of a buckled sheet viewed along the vorticity direction of the shear flow and definition of the mid-point orientation angle $\theta$ and mid-point curvature $\bar \kappa$. The compressional quadrant and the extensional quadrant are shown. In this schematic, the sheet is oriented in the compressional quadrant ($-\pi/2<\theta<0$).
\label{fig1}}
\end{center}
\end{figure}
 Mylar sheets (Young's modulus $E\simeq 4\;\rm{GPa})$ of different thicknesses ($h= 23,\ 50\ \rm{and}\ 125\;\rm{\mu m}$), width $w=1\;\rm{cm}$ and length $L$ ranging from $1$ to $4\;\rm{cm}$ were used. Corresponding sheet bending modulii are $B \simeq 5.0\times 10^{-6},\ 5.0\times 10^{-5},\ \rm{and}\ 8.1\times 10^{-4}\;\rm{J}$ ($B=E h^3 /12 (1 - \nu^2)$ where $\nu\simeq 0.5$ is the Poisson's ratio). The shear cell is composed of a belt driven by two co-rotating cylinders of diameter $6\; \rm{cm}$ -- see fig.\ref{fig1}(a). A  motor, connected to one of the cylinders, imposes a controlled shear rate in the range $\dot \gamma =  0.4- 10\;  \rm{s^{-1}}$. We considered single sheets, and pairs of parallel sheets with separation distance $d$ varying in the range $d/L=0.03-1$. The sheets were placed in glycerol (the viscosity is $\eta \simeq 1\;\rm{Pa.s}$ and the density is $\rho\simeq 1.2\times10^{3}\;\rm{kg.m^{-3}}$). The maximum Reynolds number $Re = \rho \dot \gamma L^2/\eta$ is of order 1 at the maximum shear rate. With the use of tweezers, the sheets were set with their normal vector in the flow-gradient plane. The cylinders were set in motion after the sheets were placed. During the dynamics, the normal vector remains in the flow-gradient plane and thus the dynamics is two-dimensional. Optical measurements with a camera were carried out from the top, i.e. along the vorticity direction of the undisturbed shear flow. We extracted from the images the midpoint orientation angle $\theta(t)$ by fitting a line to each sheet's profile -- see fig.\ref{fig1}(b). The mid-point curvature $\bar \kappa(t)$ was obtained by fitting a parabola to each sheet's profile. Measurements were performed with a time resolution of $0.1\;\rm{s}$ and with a spatial resolution of $25\;\rm{\mu m / pixel}$. Even though the maximum  $Re$ is of order $1$, in analyzing the results we will consider a low Reynolds approximation (Stokes flow). As it will appear later, this approximation gives a reasonable agreement between the simulations and the experimental data.
%%%%%%%%%%%
\section{Simulation Method} 
 %%%%%%%%%%%
We simulated the fluid-structure interaction of thin sheets in Stokes flow by a regularized Stokeslet approach \cite{smith2009boundary,olson2015hydrodynamic,montenegro2017microscale,Giulapof}. The regularized Stokeslet method has been used to study a variety of fluid-structure interaction problems at low Reynolds number, including cilia-driven transport \cite{smith2009boundary}, flagella synchronization \cite{olson2015hydrodynamic}, and flow around double helices \cite{montenegro2017microscale}. As in the experiment the flow and the sheet dynamics are two-dimensional, we simplified the simulation choosing a two-dimensional description. For a two-dimensional slender body, the approach consists in placing regularized force singularities along the body's centerline. The integral of the regularized force density over each discretization line segment represents the force exerted by that segment of the slender body on the fluid. Owing to the linearity of the Stokes equation, the velocity field $\mathbf{u}(\mathbf{x},t)$ at position $\mathbf{x}$ and time $t$ obeys the following boundary integral equation \cite{pozrikidis1992boundary}:
%%%%%%%
\begin{eqnarray}
\label{eq1}
\mathbf{u}(\mathbf{x},t)=\mathbf{u}^\infty(\mathbf{x})+\frac{1}{4\pi\eta}\int_{C} \mathbf{G}_{\epsilon}(\mathbf{x},\mathbf{x}_0)\cdot\mathbf{f}(\mathbf{x}_0,t)\mathrm{d}l,
\end{eqnarray} 
%%%%%%%
where $\mathbf{u}^\infty$ is the undisturbed background flow, $\eta$ is the dynamic viscosity, $\mathbf{f}(\mathbf{x}_0,t)$ is the force density exerted on the fluid by the sheet element $\mathrm{d}l$ centered at $\mathbf{x}_0$ and $\mathbf{G}_\epsilon$ is a 2D regularized Stokeslet for an unbounded flow \cite{cortez2001method}. Here, we neglected the double-layer potential because of the inextensibility approximation for the sheets \cite{montenegro2017microscale,Giulapof}. Since the sheets are inertia-less, the hydrodynamic force $(-\mathbf{f})$ is balanced by the local internal elastic force. Numerically, we compute the elastic force from the derivative of the bending energy, as done in \cite{fauci1988computational,olson2015hydrodynamic,Giulapof}.
The kinematics of each sheet is governed by the no-slip boundary condition on the surface of the sheet. In the slender body approximation, this condition is approximated by a no-slip condition at the centerline of the sheet:
\begin{eqnarray}
\label{eq3}
\frac{\partial\mathbf{X}(s,t)}{\partial t}=\mathbf{u}(\mathbf{X}(s,t)),  
\end{eqnarray}
where $\mathbf{X}(s,t)$ is the position vector along the centerline of the sheet at the curvilinear coordinate $s$ and time $t$. In this numerical method the sheet has zero thickness, therefore in order to observe tumbling and bending, the sheet needs to be initialized at an orientation angle different from $ \pm \pi/2$ and the initial shape set to a perturbation from a straight line \cite{PRLShelley01}. Based on our previous work \cite{Giulapof} we chose for the initial orientation $\theta_0= -\pi/2 +  \pi/10 $ and for the initial shape perturbation the first buckling mode $\kappa(s) = \kappa_0\sin (s \pi/L)$ with a small amplitude $ \kappa_0 = 8\times 10^{-3}/L $, where $\kappa(s)$ is the local curvature at $s$. At each time step, the velocity field is calculated first by eq.(\ref{eq1}), then eq.(\ref{eq3}) is advanced in time by a first-order explicit Euler scheme to obtain the sheet's configuration at the new time step. In the simulations, each sheet is discretized by 51 nodes and the  time step is $10^{-5} \dot\gamma^{-1}$. Validations of the code on two cases for which asymptotic solutions are known can be found in our previous article \cite{Giulapof}. Those two cases are the relaxation of an initially deformed sheet in a quiescent flow and the tumbling dynamics of a single sheet in a shear flow. During the simulation, at each time step the mid-point curvature $\bar \kappa(t)=\kappa(s=L/2,t)$ of a sheet is calculated by fitting a parabola to its center.
%%%%%
\section{ Dynamics of a single sheet}
%%%%%
For an inextensible flexible sheet of length $L$, width $w$ and bending modulus $B$, the Euler buckling force for axial compression scales proportionally to $w B/L^2$ \cite{ARFMbico18,Audoly2000ElasticityAG}. The viscous compressive force in a shear flow in the Stokes limit scales as $ \eta \dot \gamma L w$. Its dependence on the orientation angle is $- 2 \sin \theta \cos \theta $ \cite{ARFLindner,D0SM02184A}, which is maximum when the sheet is oriented along the compressional axis $\theta=-\pi/4$ of the shear flow (see fig.\ref{fig1}(b)). The buckling dynamics of a single flexible sheet depends therefore on the elasto-viscous number \cite{D0SM02184A}
%%%%%
\begin{eqnarray}
E_v = \frac{\eta \dot \gamma L^3}{B}.
\end{eqnarray}
%%%%%
This non-dimensional number can be also interpreted as the ratio of two time scales: $1/\dot \gamma$, the characteristic time scale of the shear flow, and $\eta L^3/B$, the characteristic time scale of curvature relaxation in a quiescent viscous liquid.
%%%%%
\begin{figure}
\begin{center}
\includegraphics{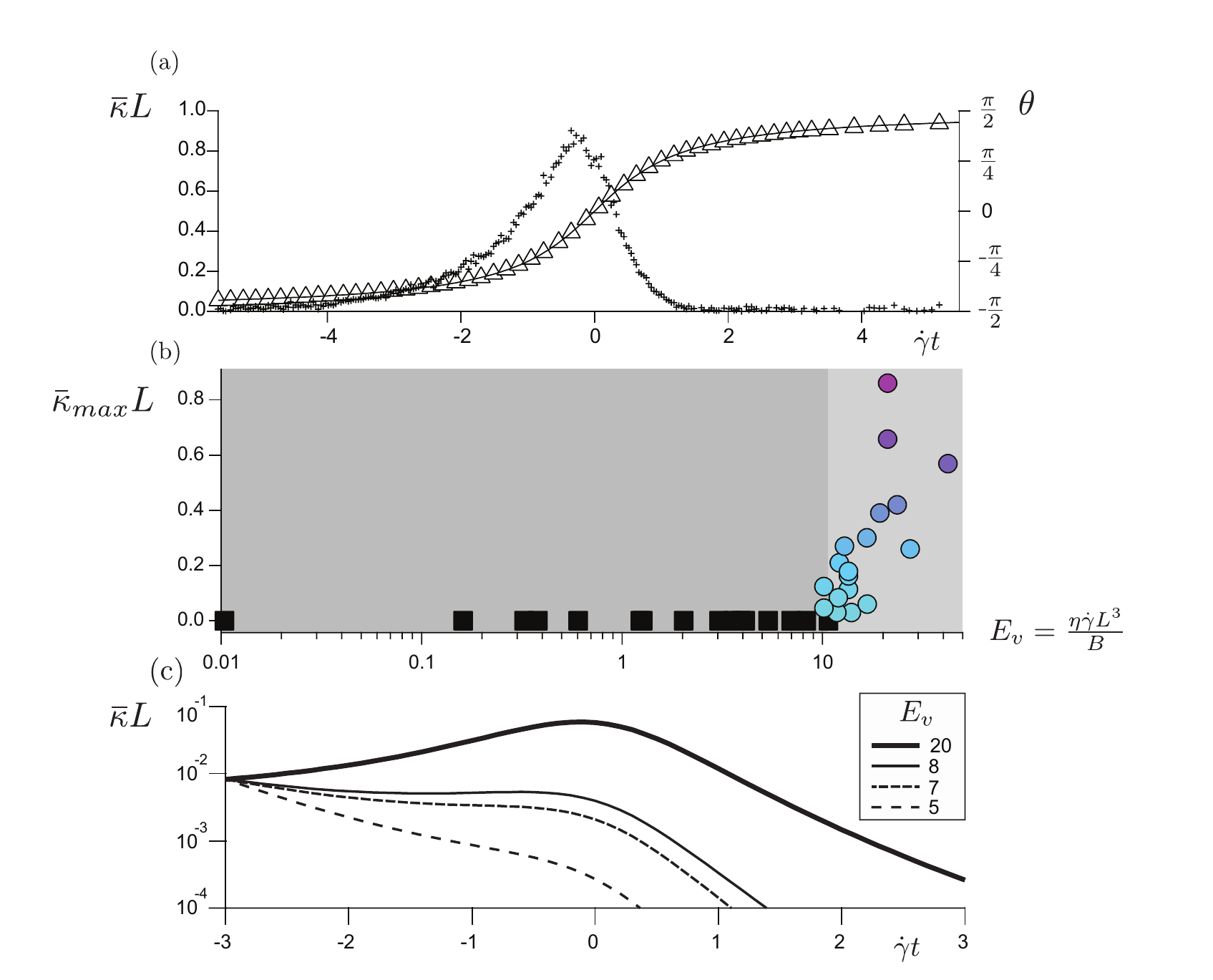}
\caption{(a) Normalized mid-point curvature $\bar \kappa L$ (crosses) and mid-point orientation angle $\theta$ (triangular markers) versus rescaled time $\dot \gamma t$ for $E_v \simeq 21$. Time has been shifted so that $\dot \gamma t = 0$ corresponds to the orientation $\theta=0$. The black line is Jeffery's prediction $\theta(t) = \arctan (\dot \gamma t)$ \cite{Jeffery}. (b) Maximum normalized curvature versus elasto-viscous number for a single sheet. The dark and light grey regions delimit the rigid limit and the buckling region, respectively. The measured critical elasto-viscous number from this diagram is $E_v^c \simeq 11$. (c) Normalized curvature versus normalized time from dynamic simulations of a single sheet for different elasto-viscous numbers.
\label{fig2}}
\end{center}
\end{figure}
%%%%%

We determined the single-sheet buckling threshold by measuring experimentally the sheet curvature $\bar \kappa$ corresponding to different elasto-viscous numbers, placing only one sheet in the shear cell. For small elasto-viscous numbers, the sheet tumbles in the flow and remains straight. For relatively large elasto-viscous numbers, for example $E_v \simeq 21$ (fig.\ref{fig2}(a)), the sheet deforms during tumbling. The time dependence of the angle $\theta(t)$ in fig.\ref{fig2}(a) is well described by Jeffery's solution for rigid oblate ellipsoids \cite{Jeffery,kamalJFM21,D0SM02184A}. This agreement validates the Stokes flow assumption we made for the simulations. This agreement also shows that the tumbling dynamics is not significantly affected by the sheets deformations for curvatures smaller than $1/L$. The time-dependent curvature is seen to grow when the sheet is oriented in the compressional quadrant ($-\pi/2\!<\!\theta\!<\!0$), which is the signature of the buckling instability. Then the curvature decays to zero, over a time scale $1/\dot{\gamma}$, as $\theta(t)$ spans the extensional quadrant ($0\!<\!\theta\!<\!\pi/2$) --~see fig.\ref{fig2}(a). To identify the single-sheet buckling threshold, we measured the maximum curvature $\bar \kappa_{max}$ attained during a tumbling cycle for different elasto-viscous numbers (see fig.\ref{fig2}(b)). The results lie in two regions sharply separated by a critical elasto-viscous number $E_v^c \simeq 11$ above which the sheet deforms with a curvature larger than the experimental resolution. Below this number the sheets curvature is negligible. To corroborate this observation, we performed numerical simulations of single sheets for different elasto-viscous numbers, see fig.\ref{fig2}(c). For elasto-viscous numbers larger than $E_v = 8$, the curvature increases in time, signature of the growth of the buckling instability. For elasto-viscous numbers smaller than $E_v = 8$, the curvature decays. The agreement between the numerical prediction ($\simeq 8$) and the experimental value ($\simeq 11$) is acceptable considering the finite experimental resolution, which makes a very precise determination of the buckling threshold difficult \cite{ARFLindner}. In literature, a mathematical model based on applying Jeffery's solution for the hydrodynamic stress on an oblate ellipsoids to a flexible disk predicted a threshold values $\simeq 10^2$ \cite{LINGARD1974119}. Recent simulations of an hexagonal flexible sheet modeled as a collection of beads interacting via long-range hydrodynamic interactions - represented at the Rotne-Prager-Yamakawa level - suggested a critical buckling threshold of about $50$ \cite{D0SM02184A}. Since the hydrodynamic compressive force depends on the shape, it is expected that the buckling threshold for rectangular sheets is different than the ones for ellipsoids and hexagonal sheets, so differences with published work are expected. The experimental determination of the buckling threshold for single rectangular sheets, confirmed by our numerical simulation, is an important step that provides a reference case for the study of pairs of parallel sheets.

\section{ Dynamics of a pair of parallel sheets}
\begin{figure}[h]
\begin{center}
\includegraphics{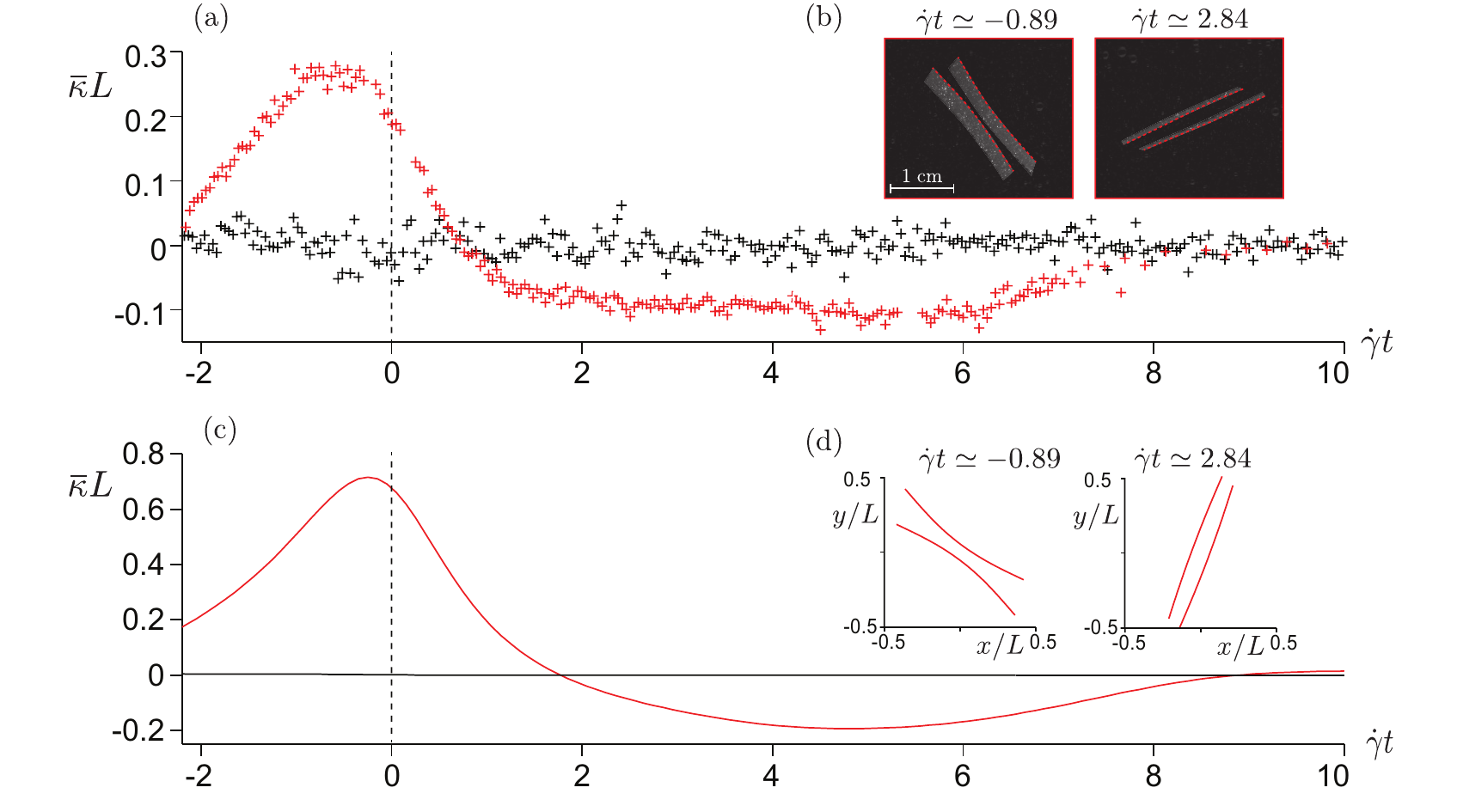}
\caption{Comparison between single sheet dynamics and dynamics of a pair of parallel sheets below the single-sheet buckling threshold. (a) Experimental normalized curvature $\bar \kappa L$ versus normalized time $\dot \gamma t$ for a single sheet (in black) and a sheet pair separated by $d/L\simeq 0.2$ (in red) for $E_v \simeq 3.6$. (b) Images of a pair of parallel sheets at two selected times. (c) Normalized curvature $\bar \kappa L$ versus normalized time $\dot \gamma t$ for simulation of a single sheet (in black) and a pair of sheets separated by $d/L= 0.1$ (in red) for $E_v = 7$. (d) Simulated shapes of a pair of parallel sheets corresponding to the two selected times of fig.\ref{fig3}(b).
\label{fig3}}
\end{center}
\end{figure}
A body formed by two sheets bonded together by adhesion or friction has a larger bending rigidity than a single sheet \cite{Poincloux21,MultiPRL19,Audoly2000ElasticityAG}. Therefore one may intuitively assume that two sheets separated by a layer of viscous liquid would have a larger buckling threshold than a single sheet. In contrast, we found that a pair of parallel sheets can deform for values of $E_v$ below the single-sheet threshold. For example, for $E_v \simeq 3.6 $ the single-sheet curvature is negligible (see fig.\ref{fig2}(b) and fig.\ref{fig3}(a)) while for the same parameter two sheets separated by $d/L \simeq 0.2$ display a finite curvature. The curvature of the two sheets increases with time, then decreases, changes sign and finally decays to zero at the end of the tumbling motion (see fig.\ref{fig3}(a)). In the single-sheet case, for $E_v >E_v^c$ the curvature relaxes while the sheet is oriented in the extensional quadrant (see fig.\ref{fig2}(a)). In contrast, pair of sheets deform while oriented in the extensional quadrant (fig.\ref{fig3}(b) right panel). These two changes of behavior for a pair of sheets, bending below the buckling threshold and bending in the extensional quadrant, are consequences of hydrodynamic interactions between the sheets, as it will be demonstrated below.

To rationalize the experimental observations, we simulated the dynamics of two parallel flexible sheets. For $E_v = 7$, the simulations indicate that the single sheet dynamics is stable: a small initial curvature decreases in time -- see the black line in fig.\ref{fig3}(c). For a pair of parallel sheets separated by a distance $d=0.1L$ and the same value $E_v=7$, the computed curvature follows qualitatively the experimental dynamics, see the red line in fig.\ref{fig3}(c): each sheet of the pair deforms, adopts a concave shape in the compressional quadrant (fig.\ref{fig3}(d), left panel), then the curvature changes sign, the sheets adopt a convex shape in the extensional quadrant (fig.\ref{fig3}(d), right panel) and finally the deformation relaxes to zero. Because in the simulation only hydrodynamic interactions are accounted for, the simulation results support the hydrodynamic origin of the two changes of behavior discussed above in relation to experiments.
\begin{figure}[h]
\begin{center}
\includegraphics{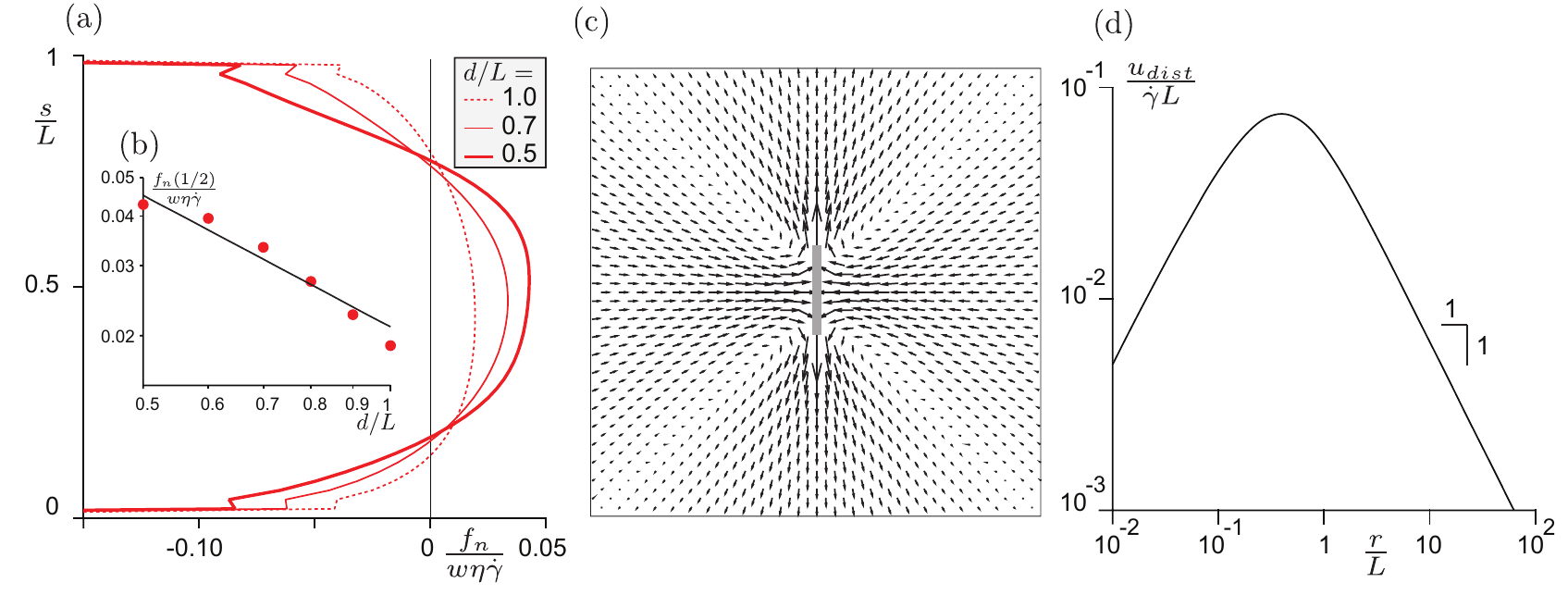}
\caption{Hydrodynamic interactions from simulation. (a) Lateral force induced by the first sheet on the second sheet in the case of two parallel sheets in a shear flow, oriented in the compressional quadrant (at $\theta_0 =- \pi/2 + \pi/10$), for $d/L = 1,\ 0.7 \ \rm{and}\ 0.5 $. (b) Magnitude of the lateral force at the center of the sheet versus the separation distance  when the sheets are oriented in the compressional quadrant (at $\theta_0 =- \pi/2 + \pi/10$). The line is the best fit $y=A x^{\alpha}$ with $A\simeq 0.02$ and $\alpha \simeq -1.1$. (c) Vector plot of the 2D disturbance flow. The rectangle represents the sheet. (d) Magnitude of the disturbance flow velocity $u_{dist}$ induced by a sheet in a 2D compressional flow, versus the distance $r$ measured orthogonally to the sheet. }
\label{fig4}
\end{center}
\end{figure}

From the numerical simulations of two sheets we computed the lateral force on one of the two sheets, when oriented in the compressional quadrant and for varying distance $d/L$, see fig.\ref{fig4}(a). The lateral force is non-uniform along the sheet, with a maximum value in the center of the sheet and minima located at the two edges. The force distribution can be described, to a first approximation, as a parabola. As $d/L$ increases it is seen from fig.\ref{fig4}(b) that the amplitude of the parabolic profile decreases, following a power law with an exponent close to $-1$. To explain and model this lateral force, we quantified the disturbance flow field set up by a sheet. Because the sheet is inertia-less and the flow 2D, the disturbance flow field in the far field is that of a 2D force dipole whose amplitude decreases as $1/r$ \cite{BookMorris}, where $r$ is the distance from the geometric center of the sheet. The flow disturbance induced by a body oriented along the compressional or extensional axis of a shear flow can be approximated by placing an elongated particle in a two-dimensional purely straining flow \cite{batchelor1971stress}, with the long axis of the particle along the extensional direction. We performed simulations with this simplified flow configuration. The computed vector plots in fig.\ref{fig4}(c) illustrate the dipolar characteristics of the flow, where it is seen that the spatial variation of the flow corresponds to the parabolic distribution of the lateral force. The amplitude of the disturbance flow field is reasonably well captured by a $1/r$ dependence for $r$ as small as $ 0.5L$ (see fig.\ref{fig4}(d)). The sign of the background straining flow governs the sign of the dipole: with our convention the sign is positive for compressional background flow and negative for extensional background flow. Hence, the simulations show that the presence of one sheet generates a parabolic lateral force on the other sheet; this lateral force originates from the disturbance dipole flow field, its amplitude scales as $L/d$ and its sign is governed by the background flow field, being positive when the sheets are in the compressional quadrant and negative in the extensional quadrant.

\begin{figure}[h!]
\begin{center}
\includegraphics{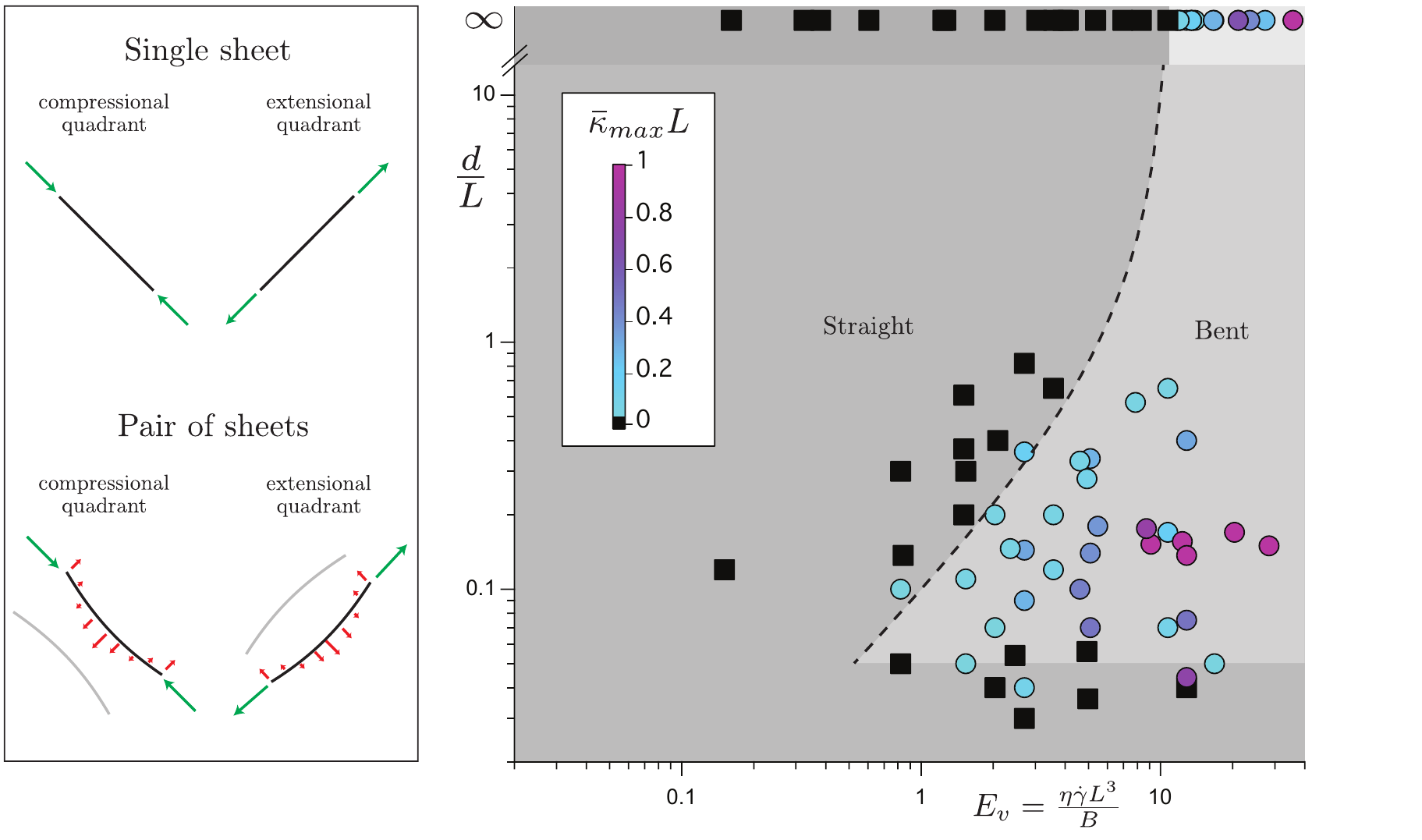}
\caption{Left panel: sketch of the hydrodynamic forces distribution in the single and two sheet cases. The black line segments represent the sheets. The green arrows represent the compressional or extensional tangential forces. The red arrows represent the dipolar lateral forces. For $E_v<E_v^c$ the single sheet remains straight while the pair of sheets adopts concave and convex shapes. Right panel: morphology diagram. Maximum normalized curvature for different normalized separation distances $d/L$ and elasto-viscous numbers $E_v = \eta \dot \gamma L^3/B$. The black squares correspond to deformation below the experimental resolution $\kappa_r= 0.01/L$, the color circles to a finite curvature. The shadow regions are guides for the eye. The experimental data for single sheets of fig.\ref{fig2}(b) are reported in correspondence to $d/L = \infty$. The equation of the dashed line is $d/L = K E_v/\left( \kappa_r L (E_v^c - E_v)\right) $, with $E_v^c = 11$, $\kappa_r L=0.01$ and $K=0.01$. The dashed line is plotted for $d/L>0.05$}
\label{fig5}
\end{center}
\end{figure}

The information above enables to construct a minimalistic model of flow-induced shape changes that takes into account the dependence on sheet-to-sheet distance. From a balance of forces and moments on an inextensible sheet, in the linear approximation the curvature $\kappa$ obeys the Euler-Bernoulli equation 
\begin{eqnarray}
B w \frac{d^2 \kappa}{ds^2}   - T_t(s) \kappa(s) - f_n(s) =  0,
\end{eqnarray}
 where $s$ is the curvilinear coordinate, $f_n$ is the lateral hydrodynamic force per unit length and $T_t$ is the axial tension \cite{Audoly2000ElasticityAG,Wexler2013aa}. The axial tension satisfies 
 \begin{eqnarray}
 \frac{dT_t}{ds} + f_t(s) = 0,
 \end{eqnarray}
where $f_t$ is the axial hydrodynamic force per unit length \cite{Audoly2000ElasticityAG,Wexler2013aa}. To model $f_n$ and $f_t$ we used a quasi-static approximation that consists of two main assumptions. First, we neglected the effect of the lateral hydrodynamic drag force caused by the time variation of the curvature. Second, we assume that the curvature is only coupled to the orientation $\theta$ through the amplitude of $f_t$, which we assume to be $-2\sin \theta \cos \theta$. Considering the two extreme cases $\theta=-\pi/4$ (orientation at maximum compression) and $\theta=\pi/4$ (orientation at maximum extension) and modeling the axial force per unit length as an edge force arising from the straining component of the imposed shear rate, we obtain $T_t = - \eta \dot \gamma L w$ for $\theta=-\pi/4$  and $T_t = \eta \dot \gamma L w$ for $\theta=\pi/4$. Fitting the results of our numerical simulations, we modeled the lateral force per unit length arising from the dipolar flow field as
 \begin{eqnarray}
 f_n(s) =  \pm \, w\, \frac{L}{d}\, \eta \dot \gamma\, K g(s)\ 
 \end{eqnarray}
 where the sign depends on whether the sheet is oriented along the compressional or the extensional axis. The function $ g(s) = \frac{1}{12}-\left(\frac{s}{L}-\frac{1}{2}\right)^2$ is a symmetric parabola of zero mean that reproduces the spatial variation of the lateral force seen in fig.\ref{fig4}(a) and $K$ is a numerical pre-factor. We estimated $K\simeq 0.4$ from the force amplitude computed at the orientation $\theta_0$, see fig.\ref{fig4}(b) (by definition $K = A\  g(1/2)  / 2 \sin \theta_0 \cos \theta_0$, where $A$ is a fitting parameter). The moment balance then reads
\begin{eqnarray}\label{eqn}
   \frac{d^2 \tilde \kappa}{d\tilde s^2} \pm E_v  \left(\tilde \kappa(\tilde s) - \frac{K}{\tilde d} g(\tilde s) \right)  &=&  0,
\end{eqnarray}
where $\tilde \kappa = \kappa L$, $\tilde s =  s/L$ and $\tilde d = d/L$. The "$+$" sign corresponds to the maximum compression ($\theta = -\pi/4$). The "$-$" sign corresponds to the maximum extension ($\theta = +\pi/4$). At maximum compression, for the single sheet case ($1/\tilde d \to 0 $) this differential equation reduces to the classical Euler-buckling equation for an edge axial load. If $E_v = \pi^2$, the Euler-buckling equation admits two solutions verifying the free end boundary conditions $ \tilde\kappa(0) =  \tilde\kappa(1) = 0$. One solution is the trivial solution $\tilde \kappa(\tilde s) = 0$ and the other is the first buckling mode $\tilde \kappa(\tilde s) =\tilde \kappa_0 \sin\left(\pi \tilde s\right)$ for a purely axial load. The value of the buckling threshold, here $\pi^2$, corresponds to a uniform axial tension. However, it can be shown that for a more realistic model of the axial hydrodynamic force $f_t(s)$, i.e. a linear variation along the sheet \cite{ARFLindner}, the axial tension is a parabola  and the threshold is reduced by only  $15\%$ with respect to $\pi^2$. Therefore the model of uniform axial tension captures the essential behavior of buckling. The value $E_v^c \simeq 11$ we measured experimentally is comparable with the prediction $\pi^2$ of this minimal model.
On the other hand, at the maximum compression, for finite $\tilde d$ eq.\eqref{eqn} admits only one solution satisfying the boundary conditions for any given value of $E_v$: 
\begin{eqnarray}\label{sol}
\tilde \kappa(\tilde s) =\frac{K}{6 \tilde d  E_v} \left(\left[ E_v (-6 (\tilde s-1) \tilde s-1)+12  \right]+\left( E_v-12\right) \cos \left(\sqrt{E_v} \tilde s\right)+(E_v-12) \tan(\sqrt{E_v}/2) \sin\left(\sqrt{E_v} \tilde s\right)\right).
\end{eqnarray}
The existence of a unique non-zero solution means that there is no buckling instability in the strict sense. Hydrodynamic interactions remove the buckling instability and the curvature has a finite bending amplitude for all values of $E_v$. To summarize, for a single sheet in pure compression there is a buckling instability while for a pair of sheets, there is no buckling instability but bending deformations do occur. Taking the limits $E_v\to0$ and $E_v - E_v^c \to 0$ with $E_v<E_v^c$, one can derive from eq.\eqref{sol} the following approximation for the maximum curvature (at $\theta=-\pi/4$) of the mid-point of the sheet: 
\begin{eqnarray}
\bar \kappa \sim + \, K \frac{L}{d} \frac{E_v}{E_v^c - E_v}.
\end{eqnarray}
Here we have indicated explicitly the sign of the lateral force, the $"+"$~sign corresponding to the compressional quadrant. In the extensional quadrant, by solving eq.\eqref{eqn}, one can show that the curvature scales as $\bar \kappa \sim - K E_v\, L/d $. As illustrated in the sketch on the left panel of fig.\ref{fig5}, the change of sign of the dipole force as the sheet tumbles explains the change from concave to convex morphologies seen in fig.\ref{fig3}(b). Equation \eqref{eqn} is linear, so the scaling in $L/d$ for the dipole amplitude determines the dependence of the bending curvature with respect to $d$.
%%%

To evaluate the ability of the model above to capture essential features of the experimental data, for each value of the parameters $(E_v,d/L)$, we measured the maximum rescaled curvature $\bar \kappa_{max} L$ during tumbling. For pair of sheets, the results indicate two regions of behavior (see right panel fig.\ref{fig5}). A first region where each sheet's curvature is lower than the experimental resolution. The label "Straight" in the figure indicates this first region. And a second region where the sheets deform significantly (indicated by the label 'Bent' in the figure). Significant deformations are seen to occur for $E_v$ as small as $0.8-1$, i.e. approximately ten times smaller than in the single-sheet case. Sheet proximity has thus a strong effect on the morphology.  Our simple model provides a criterion for which the curvature becomes larger than the experimental resolution $ \kappa_r = 0.01/L$. This criterion defines two regions in the morphology diagram delimited by the dashed line of equation $d/L = K E_v/\left( \kappa_r L (E_v^c - E_v)\right) $ in fig.\ref{fig5}. The model predicts a much larger amplitude of deformation than observed in the experiments, but a similar trend with respect to $E_v$. Indeed, the  value of $K$ used in fig.\ref{fig5}, $K=0.01$, is smaller than the one obtained from fitting the lateral force profiles of fig.\ref{fig4}(b), $K\simeq0.4$. The overestimation of the amplitude of deformation in the model likely originates from two aspects of the quasi-static approximation we used: first, we neglected the hydrodynamic drag force in the normal direction to the sheets, which delays the curvature response time; second, we assumed that the compressive force $f_t$ is constant while in the experiments the sheets tumbles and thus are not submitted to a constant force. But it can be seen accounting for the difference in the value of $K$ that the model is in reasonably good agreements with experimental data for $d/L\gtrsim0.05$, as the dashed line in this interval separates the circles from the squares symbols with the correct scaling law in $L/d$. For $d/L\lesssim0.05$, the sheets are observed to remain straight during the tumbling motion for all $E_v$ tested and the $L/d$ prediction fails. The experimental data of fig.\ref{fig5} reveal thus that the relation between the separation distance and the critical elasto-viscous number to observe significant bending is non-monotonic. Lubrication forces between two plates separated by a distance $d$ scale as $1/d^3$ for a normal displacement \cite{SRIDHAR20022547} and so are dominant at small distances over the dipolar forces. The time scale for the growth of the deformation in the case of a steady compressive force and lubrication scales as $(L/d)^3 \tau$ \cite{SRIDHAR20022547} where $\tau=\eta L^3/B$ is the elasto-viscous time scale. This time scale is much longer than the tumbling time scale $1/\dot \gamma$ for moderate $E_v = \dot \gamma \tau$ and small $d/L$. Thus, lubrication forces constrain dynamically the deformation for very small distances and moderate $E_v$.
\section{Conclusion}
In this study we measured for the first time the effective buckling threshold, which we define as the threshold to observe significant bending, for a pair of flexible sheets suspended in a viscous simple shear flow as function of the sheet-sheet distance. In experiments, we obtain a value of the critical elasto-viscous number for buckling of a single rectangular sheet of $E_v^c \simeq 11$. This number is quite close to the one we obtain form 2D simulations,  $E_v^c\simeq 8$. Our main result is the demonstration of a large reduction, by about a factor of ten, of the elasto-viscous number for which a close pair of parallel sheets bend significantly. This reduction is caused by the dipolar flow disturbance induced by one sheet. This disturbance induces a lateral force on the second sheet. With a minimal model we showed that this lateral force enhances the effect of the compressional force experienced by the pair when oriented along the compressional axis of the shear flow. Furthermore, we showed that the dipolar flow disturbance induces bending also when the pair is oriented in the extensional quadrant. Experiments and simulations suggest that the amplitude of bending is inversely proportional to the distance between the sheets. For small separations, the lubrication force prevails and limits the dynamical deformation of the sheets. The competition between the dipolar enhancement and lubrication leads to a non-monotonic relation between distance and effective buckling threshold.

In the applied context of designing macroscopic materials, for instance nanocomposites, from sheet-like nanoparticles by liquid-based methods (as ink printing, coating, polymer nano-composite processing and liquid-phase exfoliation \cite{Nicolosi1226419,C3MH00144J}), our results suggest that at finite volume fraction hydrodynamic interactions could amplify deformations induced by the shear flow. The effect could alter thermal, optical or electrical properties that are dependent on the nanoparticle shape. 
In the context of rheology, by focusing on hydrodynamic pair-interactions our results provide a first step to understand the dynamics of flexible sheets in suspension. In particular, it has been evidenced for suspensions of fibers that buckling produces normal stress differences \cite{PRLShelley01}. Hence, our results suggest that the microstructure of a suspension of sheet-like particles, including the statistics of pair-particle orientation and inter-particle distance, could have a profound influence on the rheology by affecting the instantaneous particle shape. Therefore, the microstructure of suspensions of sheet-like particles should be well-characterized in future rheological studies.

\paragraph*{ \bf Acknowledgment}
We thank D. Tam and B. Metzger for discussions. This project has received funding from the European Research Council (ERC) under the European Union's Horizon 2020 research and innovation program (Grant agreement No. 715475)

\bibliographystyle{unsrt}%{apsrev4-1}
\bibliography{ArXiv_PERRIN}% Produces the bibliography via BibTeX.

%%%%%%%%%

\end{document}